\documentclass{article}

\usepackage{microtype}
\usepackage{graphicx}
\usepackage{subfigure}
\usepackage{booktabs} 
\usepackage{listings}
\usepackage{aas_macros}
\usepackage{hyperref}

\usepackage[accepted]{icml2023}

\usepackage{amsmath}
\usepackage{amssymb}
\usepackage{mathtools}
\usepackage{amsthm}

\usepackage[capitalize,noabbrev]{cleveref}

\theoremstyle{plain}

\theoremstyle{definition}

\theoremstyle{remark}

\usepackage[textsize=tiny]{todonotes}

\icmltitlerunning{Stellar Spectra Fitting with Amortized Neural Posterior Estimation and nbi}

\begin{document}

\twocolumn[
\icmltitle{Stellar Spectra Fitting with Amortized Neural Posterior Estimation and \textit{nbi}}




\begin{icmlauthorlist}
\icmlauthor{Keming Zhang}{berkeley}
\icmlauthor{Tharindu Jayasinghe}{berkeley}
\icmlauthor{Joshua S. Bloom}{berkeley}

\end{icmlauthorlist}

\icmlaffiliation{berkeley}{Department of Astronomy, University of California, Berkeley, Berkeley, USA}

\icmlcorrespondingauthor{Keming Zhang}{kemingz@berkeley.edu}

\icmlkeywords{Machine Learning, ICML}

\vskip 0.3in
]



\printAffiliationsAndNotice{}  

\begin{abstract}
Modern surveys often deliver hundreds of thousands of stellar spectra at once, which are fit to spectral models to derive stellar parameters/labels.
Therefore, the technique of Amortized Neural Posterior Estimation (ANPE) stands out as a suitable approach, which enables the inference of large number of targets as sub-linear/constant computational costs.
Leveraging our new \textit{nbi} software package\footnote{https://github.com/kmzzhang/nbi}, we train an ANPE model for the APOGEE survey and demonstrate its efficacy on both mock and real APOGEE stellar spectra. Unique to the \textit{nbi} package is its out-of-the-box functionality on astronomical inverse problems with sequential data. As such, we have been able to acquire the trained model with minimal effort.
We introduce an effective approach to handling the measurement noise properties inherent in spectral data, which utilizes the actual uncertainties in the observed data.
This allows training data to resemble observed data, an aspect that is crucial for ANPE applications.
Given the association of spectral data properties with the observing instrument, we discuss the utility of an ANPE ``model zoo,'' where models are trained for specific instruments and distributed under the \textit{nbi} framework to facilitate real-time stellar parameter inference.
\end{abstract}

\section{Introduction}

The efficacy of Amortized Neural Posterior Estimation (ANPE) has been demonstrated for a wide range of astronomical problems, such as gravitational waves \citep{dax_real-time_2021}, galaxy SED fitting \citep{hahn_accelerated_2022,khullar_digs_2022}, binary-lens microlensing \citep{zhang_real-time_2021}, exoplanet atmosphere retrieval \citep{vasist_neural_2023}, among others.
The predicate of ANPE is that the observed data are drawn from the same distribution as the training data.
This requirement has substantially complicated the application of ANPE to  time-domain astronomy. For light-curve data extracted from the same time-domain survey, each photometric data point is taken under different conditions, and the light-curve time samplings can vastly different, with large temporal gaps, some sampled very sparsely, others densely. 

In comparison, spectroscopy surveys are much better positioned in the application of ANPE. 
Compared to light-curve datasets, all data points for any given spectra are taken at the same time under the same observing condition and with the same instrument.
For any given spectrograph, the wavelength grid/sampling has fixed options, and the noise heterogeneity across wavelength bins is strongly tied to the instrument set-up.
Thus, while the sensitivity and noise properties may be heterogeneous across the wavelength axis for the same spectrum, they are highly homogeneous across different spectra. Also considering the massive scale at which spectroscopic surveys are conducted, this makes the problem of spectral model fitting uniquely suited to be solved with ANPE.

The association of data properties with the observing spectrograph strongly indicates a program to develop ANPE models for the most popular spectrographs currently used, including APOGEE, LAMOST, GALAH, Gaia RVS, Keck/HIRES, LBT/PEPSI, Magellan/MIKE, etc., which will facilitate real-time analysis as soon as data is reduced. The training, distribution, and application of such amortized inference models require user-friendly frameworks for NPE. We have developed one such framework and open-source software named \textit{neural bayesian inference} (\textit{nbi}; \citealt{zhang2023nbi}), which appears in a separate paper in this workshop.

To summarize, the \textit{nbi} framework tackles two key issues that have greatly slowed the integration of NPE methods into our routine data analysis toolbox. The first issue dealt with \textit{nbi} is the inexact nature of Neural Posterior \textit{Estimation}, which we refer the reader to the other paper for details. The second aspect is the steep learning curve of NPE for the domain person less familiar for deep learning. To apply existing software frameworks supporting NPE such as \textit{sbi} \citep{tejero-cantero_sbi_2020} and \textit{LAMPE}\footnote{https://github.com/francois-rozet/lampe}, one has to first and foremost develop their own featurizer (a.k.a.\ embedding/compressor) network that compresses the high-dimensional observed data into a low-dimensional vector, which can be ``read in'' by the normalizing flow as the conditional. This need for customization contrasts sharply with the limited data types that astronomers regularly face, in particular sequential data such as light curves and spectra.

In our previous work of applying ANPE to binary-lens microlensing \citep{zhang_real-time_2021}, we have designed a hybrid CNN/RNN network (ResNet-GRU) for light-curve data, which is more effective than either ResNet or GRU alone. As demonstrated in the current work, this network architecture is also effective for spectral data. Therefore, the \textit{nbi} package integrates this ResNet-GRU network architecture\footnote{more network architectures will be provided in the future to facilitate applying \textit{nbi} to other data types.}, thus enabling off-the-shelf application to inference on sequential data. By applying the \textit{nbi} package, we were able to develop an ANPE model for the APOGEE survey with minimal customization efforts.

This paper serves as a proof-of-concept for the benefits of developing spectral inference models tied to particular instruments using the ANPE method and the \textit{nbi} package. We demonstrate the efficacy of our approach by training the \textit{nbi} package on simulated spectra produces using \textit{The Payne} \citep{thepayne}, with realistic noise properties from the APOGEE survey. We demonstrate satisfactory results both on simulated APOGEE spectra and real APOGEE spectra. Our approach may be expanded to multiple instruments to serve as powerful tools for the spectroscopy community.

\section{Data}

The Apache Point Observatory Galactic Evolution Experiment \citep{majewski2017} is a near-infrared (NIR) survey with a spectral resolving power of $R{\sim}22,500$. APOGEE observed stars in both the northern and southern hemispheres and its NIR spectra have been heavily used to study stellar evolution, stellar chemical abundances, and Galactic structure \citep{helmi2018,pricewhelan2018}. Here, we use data from the 17th data release. APOGEE DR17 contains ${\sim}657,000$ unique stars and ${\sim}2,660,000$ individual spectra \citep{apogeedr17}. 

To model the APOGEE spectra, we adopt \textit{The Payne} \citep{thepayne} which provides a simple multi-layer perceptron (MLP) model to interpolate between synthetic NIR spectra. \textit{The Payne} yields spectral flux predictions with median absolute errors of ${\sim}10^{-3}$ for most test stars. \textit{The Payne} model consists of 25 labels consisting of stellar parameters and elemental abundances, including $T_{\rm eff}$, $\log(g)$, $\rm [Fe/H]$, $v_{\rm micro}$, $v_{\rm macro}$, $C_{12}/C_{13}$ and $\rm [X/Fe]$ where X= C, N, O, Na, Mg, Al, Si, P, S, K, Ca, Ti, V, Cr, Mn, Fe, Co, Ni,Cu, and Ge. \textit{The Payne} outputs continuum-normalized spectral fluxes for 7214 wavelength grid points.

We generate a training set of ${\sim}10^6$ spectra using \textit{The Payne}. We sample 103,344 sets of $T_{\rm eff}$, $\log(g)$ and $C_{12}/C_{13}$ from the MIST isochrones \citep{mist} for stars with ages between 1--10 Gyr and where $T_{\rm eff}\in[3000~\rm K,8000~\rm K]$. Next, for each set of $T_{\rm eff}$, $\log(g)$ and $C_{12}/C_{13}$, we randomly sample 10 instances of the remaining labels ($v_{\rm micro}$, $v_{\rm macro}$, $\rm [Fe/H]$, $\rm [X/Fe]$) using the priors from \citet{thepayne}. For the test set, we select 995 stars from APOGEE DR17 without any data reduction flags. We continuum-normalize the observed spectra and apply a mask to select pixels for fitting using the utilities provided by \textit{The Payne} \footnote{https://github.com/tingyuansen/The\_Payne}. Each spectrum has 7214 wavelength pixels where each pixel is associated with a continuum-normalized flux, continuum-normalized flux error and a binary bitmask flag. The pixel bitmasks from the published APOGEE spectra can take a wide variety of values and is used to track problematic pixels in the observed spectra, but for simplicity, we convert this to a binary flag where GOOD=0 and BAD=1.

\begin{figure*}
    \centering
    \includegraphics[width=\textwidth]{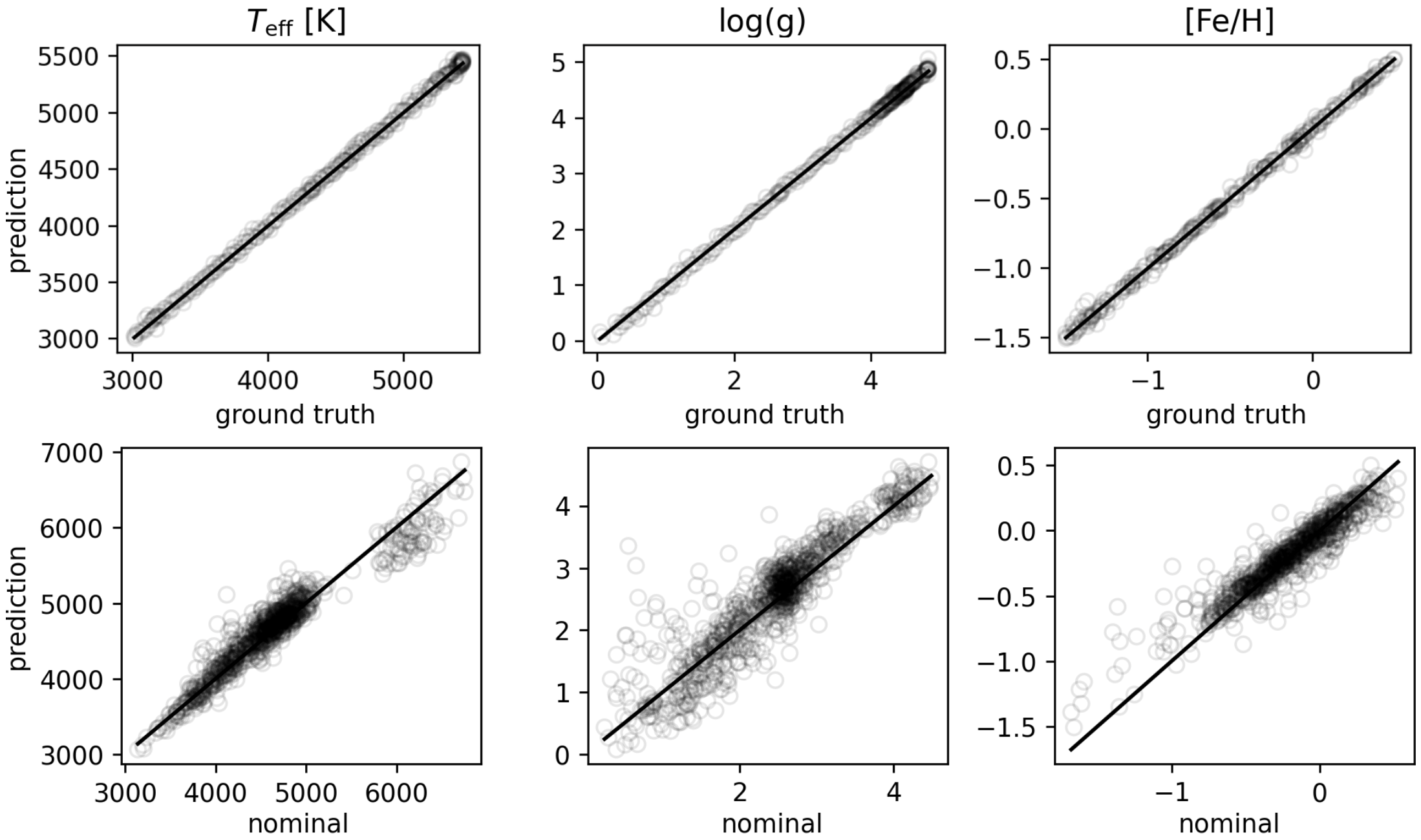}
    \caption{Top: Prediction vs.\ ground truth for a test set of simulated spectra with realistic APOGEE noise properties. Bottom: Prediction vs.\ nominal (DR 17) parameters retrieved from the APOGEE catalog, for a test set of 995 real APOGEE spectra. Note that these catalog parameters are not ground truths (see text).}
    \label{fig:label_comp}
\end{figure*}

\begin{figure*}
    \centering
    \includegraphics[width=\textwidth]{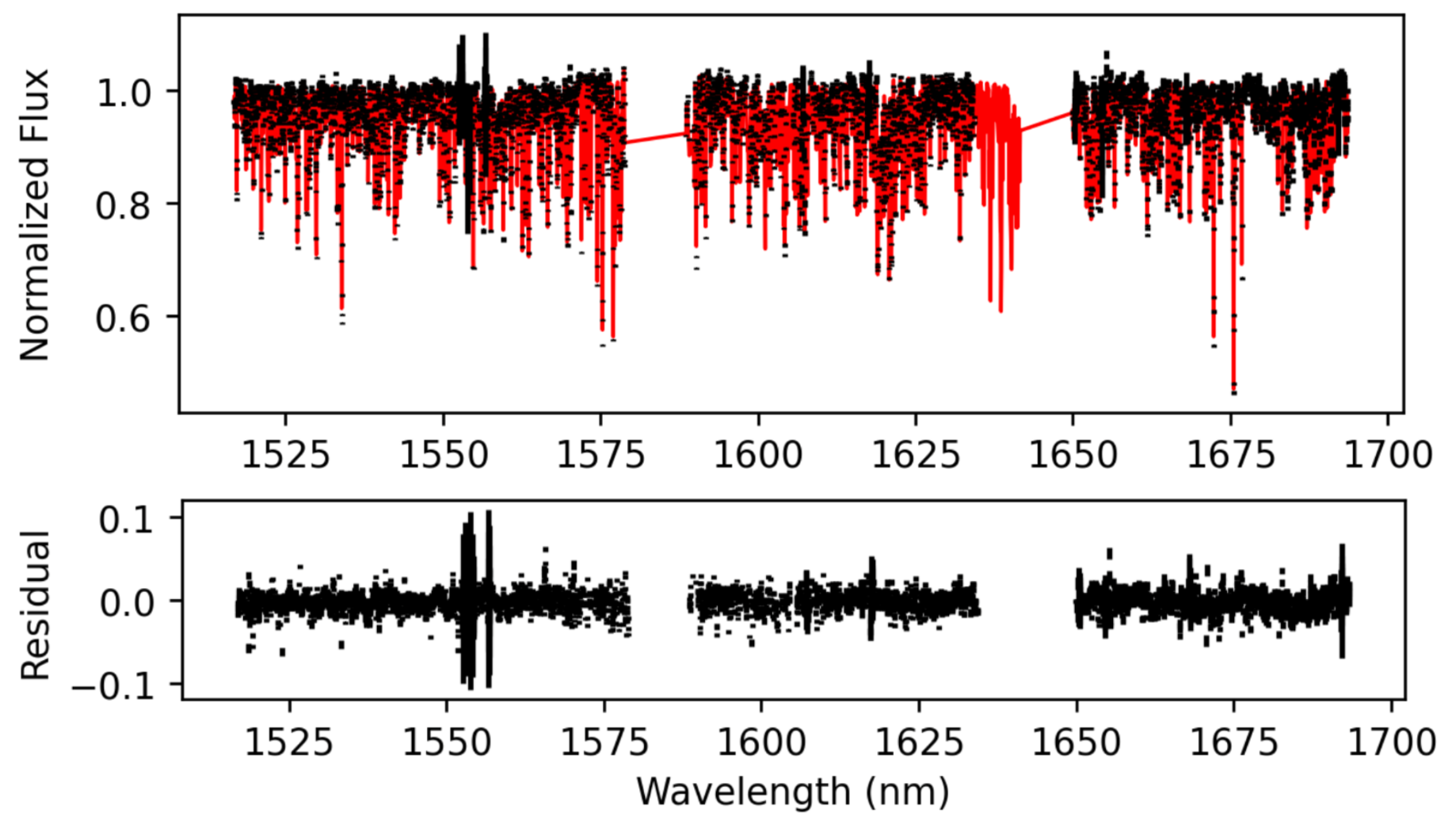}
    \caption{Example fitting of a real APOGEE spectra using \textit{nbi}. Red is the best-fit spectra model and black is observed spectra with error-bars. Masked values are not shown.}
    \label{fig:fig2}
\end{figure*}

\begin{figure*}
    \centering
    \includegraphics[width=\textwidth]{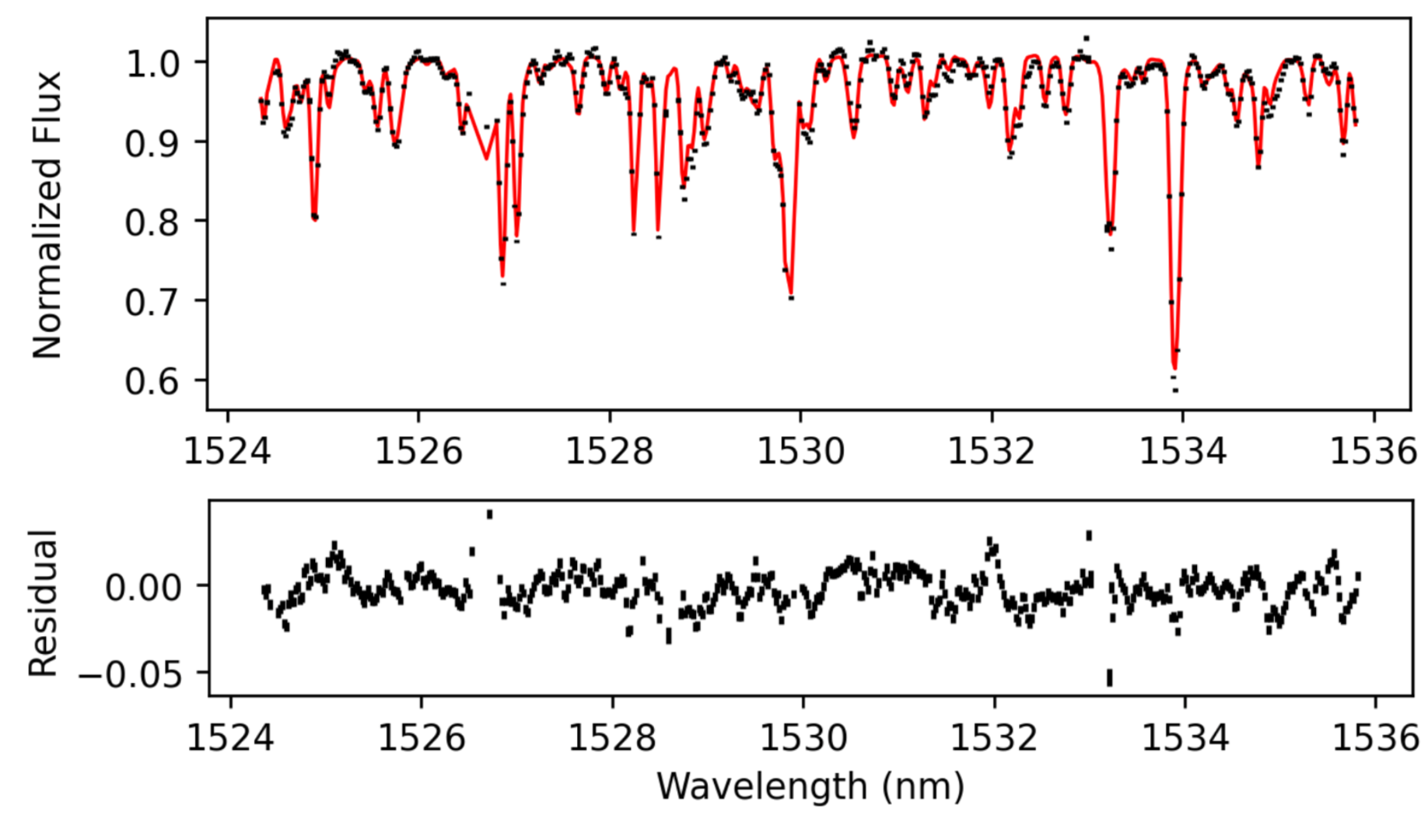}
    \caption{Same example fitting as seen in Figure 2, but zoomed in to show details.}
    \label{fig:fig3}
\end{figure*}

\section{Methods and Results}
This work exemplifies the simplicity of applying the \textit{nbi} package for inference of sequential data in astronomy. We utilize the built-in ``ResNet-GRU'' network for dimensionality reduction, with the following hyper-parameters: dim\_conv\_max=512, depth=8. We adapt the following parameters for the flow: flow\_hidden=256, num\_cond\_inputs=256, num\_blocks=15. After determining these values, the remaining required customization under \textit{nbi} is writing a function describing the noise properties, which will be called during training to apply to the noise-free simulations on the fly.

Crucial to a successful NPE application is that the training data should resemble the test data---the real spectra taken by APOGEE. Therefore, we directly use the noise (error-bars) of the our 995 test-set APOGEE spectra to be applied during training. During training, we randomly re-scale each noise sequence such that the average S/N is uniformly drawn from 25 and 600---representative of the test set. To account for the aforementioned ``bad-pixel'' bit-masks, we also adapt the 995 bit-masks of our test-set data to apply during training. Specifically, we replace each masked value with the nearest non-masked value towards the left. If the first value is masked, we then replace it with the nearest non-masked value to the right. The particular inpainting scheme does not matter, so long as it is consistent, and the aforementioned scheme is very easy to implement in practice. We also include the bit-mask as another input channel.

We then tested the trained model using both mock spectra unseen during training, as well as real APOGEE spectra. For both testing, we apply the same aforementioned masking and inpainting scheme.
In Figure \ref{fig:label_comp}, we show test results for the three most important parameters---$T_{\rm eff}$, $\log(g)$ and $\rm [Fe/H]$---out of the total of 26 parameters inferred upon. Unsurprisingly, performance on the mock test set is nearly perfectly aligned to the diagonal. Since NPE for well-specified models is a rather mature problem, we do not further discuss here, and instead focus on testing on real observed spectra.

In the lower three subplots in Figure 1, we apply the trained model to real APOGEE observed spectra and compare the \textit{nbi} predictions to the uncalibrated DR17 labels obtained from the \verb|ASPCAP| pipeline \citep{aspcap}. First, we point out that the DR17 labels are not derived using the \textit{The Payne} as the forward model. Therefore, they are not ground truths as far as this testing concerns, and the increased level of scattering is to a large extent attributed to the difference in the adopted forward model. To exemplify accuracy of the \textit{nbi} derived parameters, we simply show an example of posterior-predictive (best-fit) spectra against the actual observed spectra (Figures \ref{fig:fig2}\&\ref{fig:fig3}). The \textit{nbi} predictive spectra using \textit{The Payne} accurately captured the observed spectral features. 

Moreover, we found that over the full test set of 995 real APOGEE spectra, the fitting residuals are smaller for $95\%$ of the time for \textit{nbi}-derived parameters compared to the DR 17 APOGEE parameters. Again, we caution the difference in the adapted forward models, but this provides preliminary evidence as to the effectiveness of the NPE method and the \textit{nbi} framework for this task. We note that the [Cu/H] and [Ge/H] parameters were not provided in DR17, and we simply use the same \textit{nbi} predicted values in the above testing. However, our testing on mock spectra reveals that the \textit{nbi} model was able to provide meaningful constraint on these parameters. Therefore, the application of our trained \textit{nbi} model will be able to produce a value-added catalog when applied to the full APOGEE survey.

\section{Discussion}
In this paper, we first discuss the utility and appropriateness in training ANPE models for the problem of inference on spectral data. Due to the homogeneous noise properties for spectra taken with the instrument, ANPE is a natural fit in solving these problems. We introduce a training-time data-processing scheme to handle the noise properties, specifically by adapting the actual error-bars and bit-masks for observed spectra. The data-handling approach presented here is equally applicable to spectral fitting of galaxies, exoplanet atmospheres, etc. By applying our \textit{nbi} framework concurrently submitted to this workshop, we were able to acquire the ANPE model with minimal customization efforts.

The current work serves as a proof-of-concept for a larger program to develop ANPE models for additional spectrographs, such as LAMOST, GALAH, \textit{Gaia} RVS, Keck/HIRES, LBT/PEPSI, Magellan/MIKE, etc., which will greatly benefit analysis of such data. After the trained models are uploaded to the cloud, others may easily load and apply the model with \textit{nbi} framework. The user may directly use the trained models to acquire accurate spectral model parameters in real-time, rather than waiting $>30$ minutes in the case of commonly used tools for spectral fitting like \textit{iSpec} \citep{2014A&A...569A.111B}, providing a massive speedup.

\section*{Acknowledgment}
This work is supported by NSF award \#2206744: ``CDS\&E: Accelerating Astrophysical Insight at Scale with Likelihood-Free Inference.'' Support for TJ was provided by NASA through the NASA Hubble Fellowship grant HF2-51509 awarded by the Space Telescope Science Institute, which is operated by the Association of Universities for Research in Astronomy, Inc., for NASA, under contract NAS5-26555.




\end{document}